\documentclass[11pt]{article}
\usepackage{amsmath}
\usepackage{amssymb}
\usepackage{amsthm}
\usepackage{graphicx}
\usepackage[margin=1in]{geometry}
\usepackage{float}
\usepackage[boxed]{algorithm2e}
\usepackage{listings}
\usepackage[superscript,nomove]{cite}
\usepackage{authblk}
\usepackage{mathtools}
\usepackage[font=small,labelfont=bf]{caption}
\usepackage{xcolor}

\DeclareCaptionLabelSeparator{bar}{ \textbar  }
\SetCommentSty{mycommfont}

\DeclarePairedDelimiter{\ceil}{\lceil}{\rceil}

\title{Experimental quantum annealing: case study involving the graph isomorphism problem}

\author[1]{Kenneth M. Zick\footnote{Correspondence to \url{kzick@isi.edu}}}
\author[1,2]{Omar Shehab}
\author[1]{Matthew French}

\affil[1]{University of Southern California Information Sciences Institute (USC ISI), Arlington, Virginia, USA}
\affil[2]{Department of Computer Sciences and Electronic Engineering,
University of Maryland Baltimore County, Maryland, USA}

\date{}

\begin{document}
\SetEndCharOfAlgoLine{}

\maketitle

\begin{abstract}
\textit{
Quantum annealing is a proposed combinatorial optimization technique meant to exploit quantum mechanical effects such as tunneling and entanglement.  Real-world quantum annealing-based solvers require a combination of annealing and classical pre- and post-processing; at this early stage, little is known about how to partition and optimize the processing.  This article presents an experimental case study of quantum annealing and some of the factors involved in real-world solvers, using a \ensuremath{504}-qubit D-Wave Two machine and the graph isomorphism problem.  To illustrate the role of classical pre-processing, a compact Hamiltonian is presented that enables a reduced Ising model for each problem instance.  On random \ensuremath{N}-vertex graphs, the median number of variables is reduced from \ensuremath{N^2} to fewer than \ensuremath{N \log_2 N} and solvable graph sizes increase from \ensuremath{N = 5} to \ensuremath{N = 13}.  Additionally, a type of classical post-processing error correction is evaluated.  While the solution times are not competitive with classical approaches to graph isomorphism, the enhanced solver ultimately classified correctly every problem that was mapped to the processor and demonstrated clear advantages over the baseline approach.  The results shed some light on the nature of real-world quantum annealing and the associated hybrid classical-quantum solvers.}
\end{abstract}

Quantum annealing (QA) is a proposed combinatorial optimization technique meant to exploit quantum mechanical effects such as tunneling and entanglement \cite{mcgeoch2014adiabatic}.  Machines purportedly implementing a type of quantum annealing have recently become available \cite{boixo2014evidence}.  While the extent of “quantumness” in these implementations is not fully understood, some evidence for quantum mechanical effects playing a useful role in the processing has been appearing \cite{lanting2014entanglement, Albash2015, boixo2014computational}.  Aside from the debate over quantumness, there are interesting questions regarding how to effectively solve a real-world problem using a quantum annealer.  Quantum annealing-based solvers require a combination of annealing and classical pre- and post-processing; at this early stage, little is known about how to partition and optimize the processing.  For instance, current quantum annealers have severe practical limitations on the size of problems that can be handled.  Can the pre-processing algorithms be modified in order to improve scalability?  A second question involves post-processing.  Quantum annealers provide solutions to an “embedded” version of a problem involving physical qubits.  Post-processing is generally required for translating these to solutions to the original problem involving logical qubits (aka variables).  Occasionally, a chain of physical qubits representing a single variable resolves to an inconsistent state, a scenario known as a broken chain.  Studies are needed regarding broken chains and the possibility of classical error correction during post-processing.

This article presents an experimental case study of quantum annealing and some of the factors involved in real-world solvers, using a \ensuremath{504}-qubit D-Wave Two machine.  An example of parsimonious pre-processing is considered, along with post-processing error correction.  Through experiments on a \ensuremath{504}-qubit D-Wave Two machine, we quantify the QA success probabilities and the impact of the methods under study.  We use the graph isomorphism (GI) problem as the problem of focus. The GI problem is to determine whether two input graphs \ensuremath{G_{1,2}} are in essence the same, such that the adjacency matrices can be made identical with a relabeling of vertices.  This problem is an interesting candidate for several reasons.  First, an accurate quantum annealing-based solver for GI has never been implemented.  Second, quantum approaches can sometimes provide new insight into the structure of a problem, even if no speedup over classical approaches is achieved or even expected. Third, the GI problem is mathematically interesting; though many sub-classes of the problem can be solved in polynomial time by specialized classical solvers, the run time of the best general solution is exponential and has remained at \ensuremath{e^{O\left(\sqrt{N \log N}\right)}} since 1983 \cite{mckay2014practical, babai1983canonical}. The classical computational complexity of the problem is currently considered to be {\bf NP}-intermediate \cite{reiter2012limits}, and the quantum computational complexity of the problem is unknown.  Graph isomorphism is a non-abelian hidden subgroup problem and is not known to be easy in the quantum regime \cite{moore2008symmetric, hallgren2010limitations}.  Lastly, the GI problem is of practical interest.  It appears in fields such as very large-scale integrated circuit design, where a circuit's layout graph must be verified to be equivalent to its schematic graph \cite{kumar2009external}, and in drug discovery and bio-informatics, where a graph representing a molecular compound must be compared to an entire database, often via a GI tool that performs canonical labeling \cite{mckay2014practical}. 

This article relates to previous works as follows.  A pre-print by King and McGeoch discusses tuning of quantum annealing algorithms, including the use of low-cost classical post-processing error correction similar to what is evaluated in this article \cite{king2014algorithm}.  Our study goes further regarding pre-processing (designing a Hamiltonian to generate compact Ising models) and covers graph isomorphism rather than problems such as random not-all-equal \ensuremath{3}-{\bf SAT}.  A work by Rieffel et al. maps real-world problems such as graph coloring to a D-Wave quantum annealer \cite{rieffel2015case}.  Regarding the graph isomorphism problem in particular, multiple attempts have been made using adiabatic quantum annealing.  One of the first attempts assigned a Hamiltonian to each graph and conjectured that measurements taken during each adiabatic evolution could be used to distinguish non-isomorphic pairs \cite{hen2012solving}.  A subsequent experimental study using a D-Wave quantum annealer found that using quantum spectra in this manner was not sufficient to distinguish non-isomorphic pairs \cite{vinci2014hearing}.  A second approach converted a GI problem to a combinatorial optimization problem whose non-negative cost function has a minimum of zero only for an isomorphic pair. The approach required \ensuremath{N \lceil \log_2 N \rceil} problem variables and additional ancillary variables.  It was numerically simulated up to \ensuremath{N = 7} but not validated on a quantum annealing processor \cite{PhysRevA.89.022342}.  An alternative GI Hamiltonian was proposed by Lucas \cite{LucasIsing}.

\section*{Proposed Solver}
{\bf Preliminaries.} Before proceeding, we briefly define some key concepts.  In quantum annealing, a system is first initialized to a superposition of all possible states as dictated by Hamiltonian \ensuremath{H_{init}}, then slowly evolved such that the initializing function becomes weaker and the energy function of interest, defined by the problem Hamiltonian \ensuremath{H_P}, becomes dominant.  The time-dependent combination of the two Hamiltonians is referred to as the system Hamiltonian \ensuremath{H\left(t\right)} \cite{boixo2013experimental}:  

\begin{align}
H \left(t\right) &= A \left(t\right) H_{init} + B\left(t\right) H_P, t \in \left[0, T\right],
\end{align}
where \ensuremath{A \left(t\right)} and \ensuremath{B\left(t\right)} are monotonic functions representing the annealing schedule up to time \ensuremath{T}.  In this work, we focus on problem Hamiltonians \ensuremath{H_P}.  The problem Hamiltonian represents an Ising model, where spin variable \ensuremath{s_i \in \left\{-1,1\right\}} are subject to local fields \ensuremath{h_i} and pairwise interactions with coupling strengths \ensuremath{J_{ij}} \cite{choi2008minor}:

\begin{align}
H_P &= \sum_i h_i s_i + \sum_{i, j} J_{ij} s_i s_j.
\end{align}

It is often the case that an Ising problem \ensuremath{\left[\mathbf{h}, \mathbf{J}\right]} involves variable interactions not supported by a quantum annealing processor graph.  An example is a variable with degree higher than the maximum supported by the D-Wave Chimera architecture (six).  One common strategy is to find a {\it minor embedding} of the problem graph in the processor graph \cite{choi2008minor}, in which a set of physical qubits is used to represent each variable.  Each qubit is strongly coupled to at least one other qubit in the set, in an effort to keep the entire set in a consistent state.  These sets are commonly referred to as chains.  We refer to a minor-embedded version of a problem as an embedded Ising problem \ensuremath{\left[\mathbf{h'}, \mathbf{J'}\right]}.

{\bf Baseline Hamiltonian.} We first describe a baseline penalty Hamiltonian for the GI problem, building upon the Hamiltonian described in Lucas \cite{LucasIsing}.  The problem input is a pair of simple, undirected \ensuremath{N}-vertex graphs \ensuremath{G_{1,2}}.  Penalties are applied such that the ground state energy is zero if the pair is isomorphic and greater than zero otherwise.  The intent is for an energy minimization process (such as quantum annealing) to provide a solution to this decision problem.

A binary variable \ensuremath{x_{u,i}} is created for every possible mapping of a vertex \ensuremath{u} in \ensuremath{G_2} to a vertex \ensuremath{i} in \ensuremath{G_1}; in a solution, the variable is \ensuremath{1} if \ensuremath{u} is mapped to \ensuremath{i}, and \ensuremath{0} otherwise.  Since the model is restricted to pairwise interactions over binary variables, it represents a quadratic unconstrained binary optimization (QUBO) formulation, which can be readily converted to an Ising model.  There are two types of penalties that can be applied to an interaction between two variables.  One type (\ensuremath{C_1}) penalizes vertex set mappings that are not bijective, for instance a mapping in which both \ensuremath{x_{1,1}} and \ensuremath{x_{2,1}} are set to \ensuremath{1} (since vertices \ensuremath{1} and \ensuremath{2} in \ensuremath{G_2} cannot both map to vertex \ensuremath{1} in \ensuremath{G_1}).  The term in the QUBO matrix corresponding to the \ensuremath{x_{1,1} x_{2,1}} interaction is thus set to \ensuremath{C_1}.  Note that as long as at least one of the two variables resolves to \ensuremath{0}, the energy of the coupling (\ensuremath{C_1 x_{1,1} x_{2,1}}) is zero.  The second penalty (\ensuremath{C_2}) applies to edge inconsistencies.  An example of an edge inconsistency is when vertices \ensuremath{u,v} are mapped to \ensuremath{i,j}, and \ensuremath{u v \in E_2} while \ensuremath{i j \not\in E_1}.  Additional details regarding this style of Hamiltonian can be found in Lucas \cite{LucasIsing}.

In Lucas \cite{LucasIsing}, couplings can incur either zero, one, or two penalties.  Here, we require that couplings be penalized no more than once, in order to achieve a simple set of coupler values (e.g. \ensuremath{\left\{0, 1\right\}} instead of \ensuremath{\left\{0, 1, 2\right\}}) amenable to quantum annealing.  This is enforced by setting \ensuremath{C_1 = C_2 > 0}, and preventing double penalties.  An edge-related penalty is applied to a coupling only if there is not a vertex mapping penalty.  As an example, if vertices \ensuremath{u,v} are mapped to \ensuremath{i,j} and \ensuremath{i = j}, then the bijection has been violated and coupling \ensuremath{x_{u,i} x_{v,j}} will incur a vertex mapping penalty (\ensuremath{C_1}); therefore, an additional edge inconsistency penalty (\ensuremath{C_2}) need not be applied.  Edge inconsistency penalties are only applied if \ensuremath{i \ne j} and \ensuremath{u \ne v}.  The complete baseline Hamiltonian is 

\begin{align}
\label{eq:lucas-modified}
H_1 &= C_1 \sum_u \left(1 - \sum_i x_{u, i}\right)^2 + C_1 \sum_i\left(1 - \sum_u x_{u, i}\right)^2\nonumber \\
& + C_2 \sum_{\substack{i,j \not\in E_1\\
i \ne j}} \text{\hspace{0.1cm}} \sum_{u, v \in E_2} x_{u, i} x_{v, j} + C_2 \sum_{i,j \in E_1}  \text{\hspace{0.1cm}} \sum_{\substack{u, v \not\in E_2\\
u \ne v}} x_{u, i} x_{v, j},
\end{align}
where the \ensuremath{i \ne j} and \ensuremath{u \ne v} conditions represent the main modifications to Lucas \cite{LucasIsing}. This QUBO form can be converted to an Ising form using the relation \ensuremath{x_{u, i} = \frac{S_{u, i} + 1}{2}}.

{\bf Compact Hamiltonian.} The approach embodied in the baseline Hamiltonian \ensuremath{H_1} suffers from a severe lack of scalability.  For \ensuremath{N}-vertex input graphs, it requires \ensuremath{N^2} logical variables.  Moreover, due to the limited direct connections between qubits in the D-Wave Chimera architecture, problems are often given a minor embedding into the processor working graph.  This typically involves replicating variables across multiple qubits.  Thus, the qubit requirements can reach \ensuremath{O\left(N^4\right)}.  Problems mapped in this way to a \ensuremath{\sim500}-qubit processor tend to be limited to \ensuremath{N = 5} or \ensuremath{6}.  We now investigate whether a more effective Hamiltonian can be designed.  The idea is that many variables and interactions are unnecessary, and information indicating so can be leveraged up front during the requisite pre-processing.  Note that an isomorphic mapping requires the vertices in each matched pair to have the same degree.  Thus, degree information can be used to decide whether two vertices are eligible to be matched.  We propose a compact Hamiltonian \ensuremath{H_2} that avoids creating variables for vertices of different degree.  A second, minor simplification deals with isolated vertices (\ensuremath{degree = 0}).  If \ensuremath{G_{1,2}} each have \ensuremath{k} isolated vertices, an isomorphic mapping of such vertices is trivial and thus no variables or penalties for those vertices need be modeled.  If \ensuremath{G_{1,2}} have a different number of isolated vertices, then they also have a different number of non-isolated vertices and existing variables and penalties for those will suffice.  Thus we only create variables and penalties for vertices with degree greater than zero.  

\begin{figure}[H]
\centering
    \includegraphics[scale=0.5,clip=true,trim=0 15 0 0]{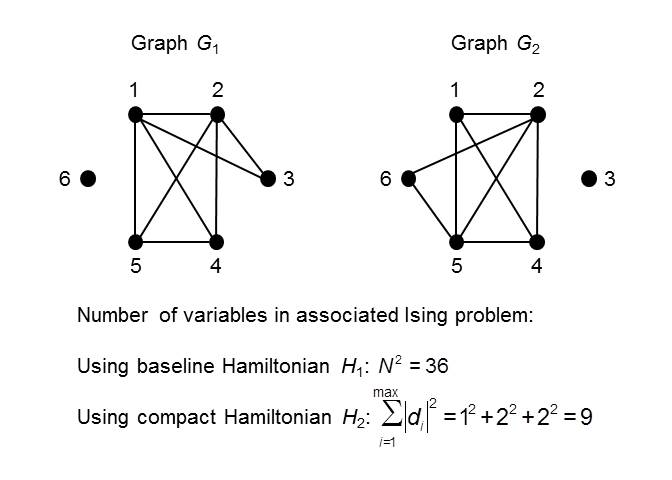}
    \caption{{\bf Example of variable reduction.}}
    \label{fig:figure1}
\end{figure}

Given the two enhancements, the total number of variables required is \ensuremath{\sum^{max}_{i=1} |d_i|^2} where \ensuremath{|d_i|} is the multiplicity of the set \ensuremath{d_i} containing vertices of degree \ensuremath{i}.  In the worst case of regular graphs of degree \ensuremath{r} greater than zero, all nodes have the same non-zero degree and thus the simplifications provide no benefit---\ensuremath{N^2} variables are still required.  Many real-world graphs are not regular and for these the benefits can be large.  An example illustrating the concept is provided in Fig.\ ~\ref{fig:figure1}.

The entire compact Hamiltonian \ensuremath{H_2} is

\begin{align}
\label{eq:degree-based-zick}
& \forall \left( deg(i) = deg(u) > 0; deg(v) = deg(j) > 0\right)  :
\nonumber \\
H_2 &=   \sum_u \left(1- \sum_i x_{u,i}\right)^2 +  \sum_i \left(1- \sum_u x_{u,i}\right)^2 
\nonumber \\
&+  \sum_{\substack{i, j \notin E_1\\
i \ne j}} \text{\hspace{0.1cm}} \sum_{u, v \in E_2} x_{u,i} x_{v,j} +  \sum_{i, j \in E_1} \text{\hspace{0.1cm}} \sum_{\substack{u, v \notin E_2\\
u \ne v}} x_{u,i} x_{v,j}.
\end{align}

\ensuremath{H_2} was validated using a software solver (D-Wave's ising-heuristic version 1.5.2) that provides exact results for problems with low tree width.  In exhaustive testing of all \ensuremath{2^{12}} \ensuremath{N = 4} pairs and \ensuremath{2^{20}} \ensuremath{N = 5} pairs, the ground state energy of \ensuremath{H_2} was confirmed to be zero for isomorphic cases and greater than zero for non-isomorphic.  

As a part of a quantum annealing-based solver, an algorithm can be employed that accepts a graph pair as input and uses the Hamiltonian to generate an associated QUBO problem (later converted to an Ising problem).  The proposed algorithm using the compact Hamiltonian \ensuremath{H_2} is presented (in pseudo code) in Algorithm ~\ref{tab:algorithm}.
\hfill \break

\begin{algorithm}[H]
 \KwIn{adjacency matrices \ensuremath{G_{1,2}}}
 \KwOut{QUBO matrix Q}
 \tcp{Define QUBO variables}
 \For{each node \ensuremath{v} in \ensuremath{G_2}}{
 \For{each node \ensuremath{i} in \ensuremath{G_1}}{
 \If{\ensuremath{deg(v) == deg(i)} and \ensuremath{deg(i) > 0}}{
create QUBO variable \ensuremath{x_{v,i}}
   }
 }
 }
\tcp{ Populate QUBO matrix with penalty terms} 
  \For{each pair of different QUBO variables \ensuremath{x_{u,i}} and \ensuremath{x_{v,j}}}{
    \tcp{ Assign penalty for node mapping conflict} 
    \If{i == j }{
penalize mapping \ensuremath{u\rightarrow i}, \ensuremath{v\rightarrow j}     \tcp*{Two nodes in \ensuremath{G_2} map to the same node in \ensuremath{G_1}}
   }
   \ElseIf{u == v}{
    penalize mapping \ensuremath{u\rightarrow i}, \ensuremath{v\rightarrow j}  \tcp*{ A node in \ensuremath{G_2} maps to two nodes in \ensuremath{G_1}}}

    \tcp{Assign penalty for edge discrepancy} 
    \ElseIf{\ensuremath{G_1(i, j)} != \ensuremath{G_2(u, v)}}{
    penalize mapping \ensuremath{u\rightarrow i}, \ensuremath{v\rightarrow j}      \tcp*{ Edge in one graph, non-edge in other} 
    }
  }
  \For{each QUBO variable}{
    Assign value along diagonal of Q
  }
 \caption{{\bf High-level algorithm for generating a QUBO problem using the proposed compact Hamiltonian \ensuremath{H_2}.}}
 \label{tab:algorithm}
\end{algorithm}

\begin{figure}[H]
\centering
    \includegraphics[scale=0.5,clip=true,trim=0 30 0 0]{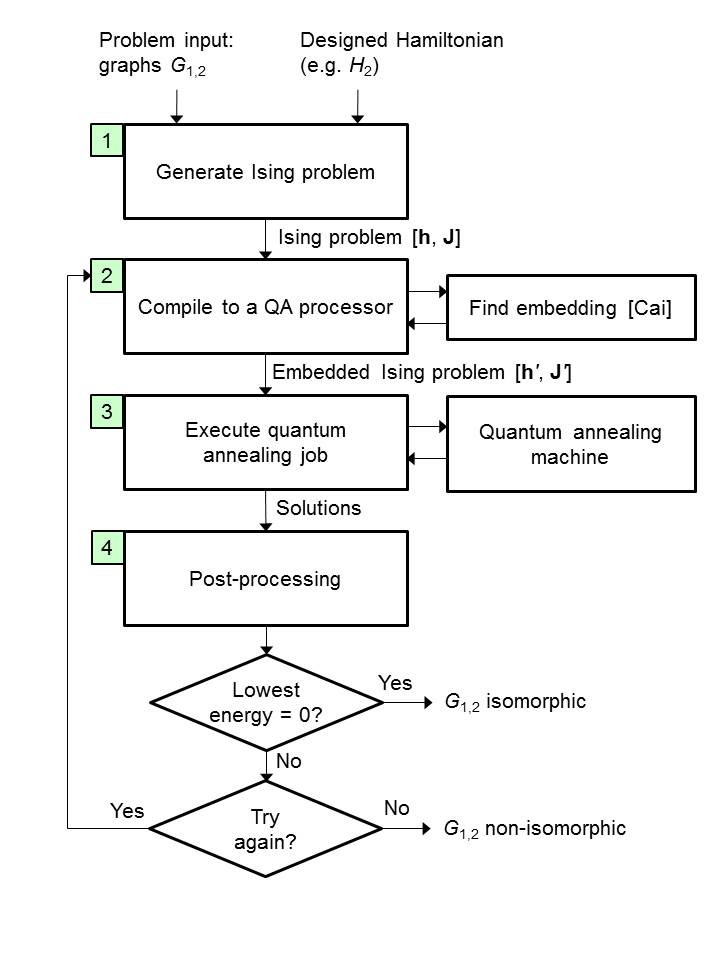}
    \caption{{\bf Graph isomorphism solver flow.}}
    \label{fig:figure2}
\end{figure}

\hfill \break
{\bf Complete Solver Flow.} The high-level flow of the proposed GI solver is shown in Fig.\ ~\ref{fig:figure2}.  The problem input is a graph pair \ensuremath{G_{1,2}}.  In this article, the graph types considered are random, simple, undirected graphs.  Graphs are generated using the Erd\H{o}s-R\'{e}nyi model \cite{erd6s1960evolution} \ensuremath{G\left(N, p\right)} where we set the probability of an edge being present \ensuremath{p = 0.5}.  An advantage of \ensuremath{G(N, 0.5)} graphs for an initial study is that all graphs are equally likely.  This type has been used in classical graph isomorphism work as well \cite{mckay2014practical}.  In step 1, the input graphs and the Hamiltonian formulation of interest (e.g. \ensuremath{H_1} or \ensuremath{H_2}) are used to generate a QUBO problem which is then converted to an Ising problem \ensuremath{\left[\mathbf{h}, \mathbf{J}\right]}.  An example of an algorithm for generating the QUBO problem is shown in Algorithm ~\ref{tab:algorithm}.  The Ising problem is then compiled to a specific quantum annealing processor in step 2.  A main task is to find sets of physical qubits to represent the problem variables (aka logical qubits); this is achieved by providing the \ensuremath{\mathbf{J}} matrix and the processor working graph to the D-Wave \texttt{findEmbedding()} heuristic \cite{cai2014practical}. Subsequently, the parameters of the embedded Ising problem \ensuremath{\left[\mathbf{h'}, \mathbf{J'}\right]} are set following certain strategies such as the use of a random spin gauge (see Methods).  The embedded Ising problem, sometimes referred to as a machine instruction, is submitted to the quantum annealing machine along with several job parameters.  The quantum annealing job is executed in step 3 and solutions are returned in the form of strings of two-valued variables.  These solutions and energies are associated with the embedded problem, not the original Ising problem.  Therefore a post-processing step is necessary (step 4), in which the state of each qubit chain is plugged into the cost function of the Ising problem.  A difficulty arises when the states of the qubits in a chain are inconsistent, a case referred to as a broken chain.  In the proposed solver, broken chains can be handled by either discarding the associated solution, or by performing majority voting over each chain.  The two strategies are compared empirically in Results.  Given a solution to the original Ising problem, the solution energy can be calculated.  If the lowest energy is zero, then the input pair can be declared isomorphic and no further jobs are necessary.  Otherwise, a decision must be made whether to repeat the process from step 2 or to stop and declare that isomorphism could not be established.

\section*{Results}
\begin{figure}[H]
\centering
    \includegraphics[scale=0.7]{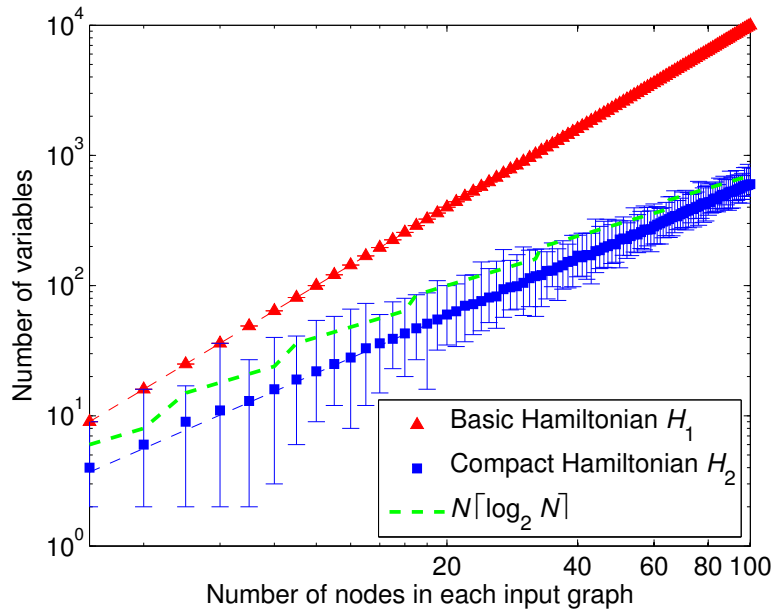}
    \caption{{\bf Scaling of the number of Ising model variables.} \ensuremath{H_1} requires \ensuremath{N^2} variables by construction; \ensuremath{H_2} scales more efficiently---the median variable requirement fits to \ensuremath{y = 0.748 N^{1.45} (R^2 = 0.9995)}.  Bars indicate the maximum and minimum.  For reference, the green dotted line represents \ensuremath{N \lceil \log_2 N \rceil}. Inputs are \ensuremath{50} isomorphic and \ensuremath{50} non-isomorphic pairs of \ensuremath{G(N, 0.5)} graphs.  }
    \label{fig:figure3}
\end{figure}

{\bf Ising Model Scaling.} To compare the resource requirements of the two proposed Hamiltonians, \ensuremath{100} pairs of graphs are used as inputs to Step 1 of the solver flow (Fig.\ ~\ref{fig:figure2}), where \ensuremath{50} pairs are isomorphic and \ensuremath{50} are non-isomorphic for each size up to \ensuremath{N = 100}.  Since \ensuremath{H_1} models a variable for each possible vertex pair, \ensuremath{N^2} variables are required by definition.  Ising problems generated using \ensuremath{H_2} are found to use fewer variables than \ensuremath{H_1}; scaling of the median problems fits to \ensuremath{0.748 N^{1.45}}. Incidentally, this indicates that most problems have fewer variables than with the Gaitan et al. approach, which entails \ensuremath{N \lceil \log_2 N \rceil} plus ancillary variables \cite{PhysRevA.89.022342}. The variable scaling is illustrated in Fig.\ ~\ref{fig:figure3}.  In addition to the number of variables, a second resource metric is the number of non-zero interactions between variables; dense interactions make the minor embedding problem more difficult.  We find that the scaling of variable interactions has been improved from \ensuremath{O\left(N^4\right)} for \ensuremath{H_1} to \ensuremath{O\left(N^{2.9}\right)} for \ensuremath{H_2} (where \ensuremath{R^2 = 0.9991}).

\begin{figure}[H]
\centering
    \includegraphics[scale=0.8]{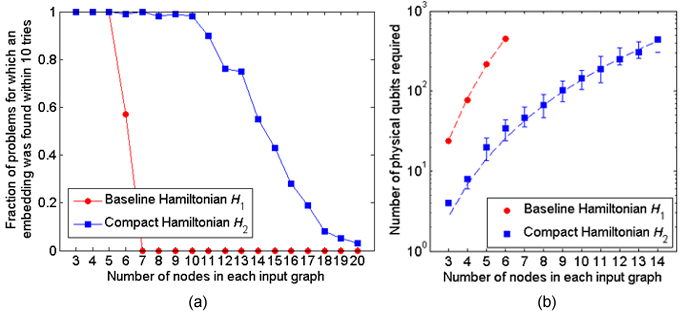}
    \caption{{\bf Embeddability when targeting the USC-LM Vesuvius processor's \ensuremath{504}-qubit, \ensuremath{1427}-coupler working graph.}  The inputs are \ensuremath{50} isomorphic and \ensuremath{50} non-isomorphic \ensuremath{G(N, 0.5)} pairs at each size.  (a) Fraction of problems for which an embedding was found within \ensuremath{10} tries, using the D-Wave \texttt{findEmbedding()} heuristic.  (b) Median number of utilized qubits among all \ensuremath{100} problems.  \ensuremath{H_1} data points fit to \ensuremath{y = 0.235 N^{4.22}} (dotted line; \ensuremath{R^2 = 0.9996}); \ensuremath{H_2} fits to \ensuremath{0.0723 N^{3.29} (R^2 = 0.9955)}.  The \ensuremath{65}\textsuperscript{th} and \ensuremath{35}\textsuperscript{th} percentiles are marked with bars (not always visible).}
    \label{fig:figure4}
\end{figure}

{\bf Embeddability.} Next, we compare the embeddability of the two approaches, in other words the extent to which Ising problems can be minor-embedded in a given processor graph.  The processor of choice is the D-Wave Two Vesuvius-6 processor housed at USC ISI.  At the time of this writing, the working graph contains \ensuremath{504} qubits and \ensuremath{1427} couplers.  Embedding is attempted using the D-Wave \texttt{findEmbedding()} heuristic \cite{cai2014practical} with default parameter values such as \ensuremath{10} “tries” per function call.  As shown in Fig.\ ~\ref{fig:figure4}a, embeddings are found for the majority of problems only for sizes \ensuremath{N \le 6} when using \ensuremath{H_1}, but sizes \ensuremath{N \le 14} with \ensuremath{H_2} (Fig.\ ~\ref{fig:figure4}a).  The median number of qubits across all problems scales as \ensuremath{O\left(N^{4.22}\right)} for \ensuremath{H_1} and has been reduced to \ensuremath{O\left(N^{3.29}\right)} for \ensuremath{H_2} (Fig.\ ~\ref{fig:figure4}b).

{\bf Experimental Quantum Annealing for Graph Isomorphism.} The accuracy of the solver described in the previous section was measured via trials conducted on a D-Wave Two Vesuvius quantum annealing processor.  Several alternative strategies were compared---the use of Hamiltonians \ensuremath{H_1} vs.\ \ensuremath{H_2}, running a single job per problem vs.\ multiple jobs, and the use of error correction.  Note that by construction of the Ising models using a penalty Hamiltonian, problems with non-isomorphic input graphs cannot achieve a zero energy state, regardless of annealing results.  The main challenge for the solver is to find the zero energy state for isomorphic pairs.  Thus, we first focus on the isomorphic case.  One hundred isomorphic pairs were input into the solver for each size \ensuremath{N} from \ensuremath{3} to \ensuremath{20}.  

\begin{figure}[H]
\centering
    \includegraphics[scale=0.75]{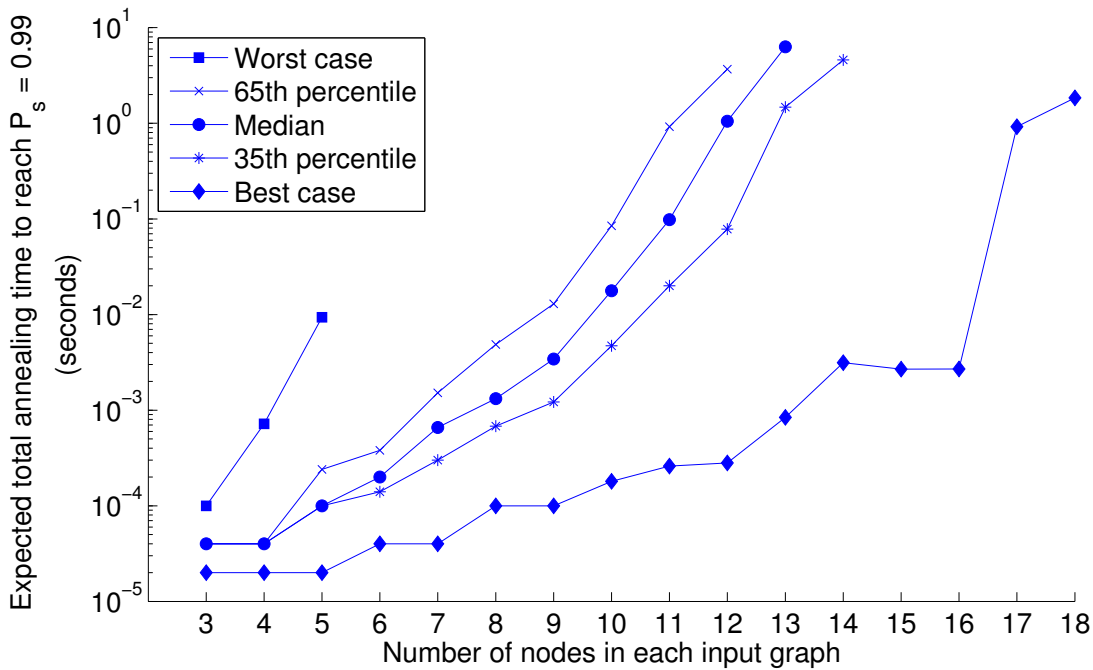}
    \caption{{\bf Total expected annealing time when using Hamiltonian \ensuremath{H_2}, multiple jobs, and classical majority voting.}  Based on the quantum annealing results for \ensuremath{100} isomorphic-input problems.  Missing data points indicate that no embedding was found and quantum annealing was not attempted. }
    \label{fig:figure5}
\end{figure}

For one strategy in particular the zero energy state was always eventually achieved---the use of Hamiltonian \ensuremath{H_2} combined with multiple jobs and error correction.  Thus, with this strategy there were no false negatives and classification accuracy reached \ensuremath{100\%} of the embeddable problems, as shown in Table ~\ref{tab:tab2}.  For the most difficult problem, the zero energy state was achieved on the \ensuremath{9}\textsuperscript{th} job.  All other strategies incurred false negatives.  For the successful strategy, the expected total annealing time was calculated (as described in Methods).  Results are shown in Fig.\ \ref{fig:figure5}.

\begin{table}[H]
\footnotesize
\caption {{\bf Number of isomorphic-input problems embedded and correctly classified as isomorphic via quantum annealing.}  One hundred problems were attempted at each problem size.  All embedded problems were solved when using \ensuremath{H_2}, error correction, and multiple jobs.}
\label{tab:tab2}
\begin{tabular}{| p{3.5cm} | l | l | l | l | l | l | l | l | l | l | l | l | l | l | l | l |}
\hline
& \multicolumn{16}{|c|} {{\bf Size of input graphs (number of vertices)}   }  \\
\hline
                                                                    & 3   & 4   & 5   & 6  & 7  & 8  & 9  & 10 & 11 & 12 & 13 & 14 & 15 & 16 & 17 & 18 \\
\hline
\hline
Number of problems for which an embedding was found using \ensuremath{H_1}        & 100 & 100 & 100 & 74 & 0  & 0  & 0  & 0  & 0  & 0  & 0  & 0  & 0  & 0  & 0  & 0  \\
\hline
Number of problems for which an embedding was found using \ensuremath{H_2}        & 100 & 100 & 100 & 99 & 99 & 96 & 95 & 93 & 87 & 83 & 54 & 39 & 20 & 7  & 4  & 1  \\
\hline
Number of problems solved using \ensuremath{H_1}, no error correction             & 100 & 99  & 98  & 13 & 0  & 0  & 0  & 0  & 0  & 0  & 0  & 0  & 0  & 0  & 0  & 0  \\
\hline
Number of problems solved using \ensuremath{H_1}, error correction                & 100 & 100 & 99  & 43 & 0  & 0  & 0  & 0  & 0  & 0  & 0  & 0  & 0  & 0  & 0  & 0  \\
\hline
Number of problems solved using \ensuremath{H_2}, no error correction             & 100 & 100 & 100 & 98 & 96 & 93 & 86 & 75 & 67 & 53 & 34 & 23 & 9  & 1  & 2  & 0  \\
\hline
Number of problems solved using \ensuremath{H_2}, error correction                & 100 & 100 & 100 & 99 & 99 & 96 & 90 & 80 & 73 & 65 & 40 & 25 & 12 & 3  & 2  & 1  \\
\hline
Number of problems solved using \ensuremath{H_2}, error correction, multiple jobs & 100 & 100 & 100 & 99 & 99 & 96 & 95 & 93 & 87 & 83 & 54 & 39 & 20 & 7  & 4  & 1 \\
\hline
\end{tabular}
\end{table}

For completeness, non-isomorphic pairs were run as well, using Hamiltonian \ensuremath{H_2} and error correction.  Since in the worst case nine jobs were required to correctly classify the isomorphic pairs above, nine jobs were submitted for each non-isomorphic problem.  One hundred non-isomorphic \ensuremath{G(N, 0.5)} problems were tested at each size between \ensuremath{N = 3} to \ensuremath{14}; of the \ensuremath{1200} problems, embeddings were found for \ensuremath{1186}.  In addition, pairs of isospectral non-isomorphic graphs (PINGs) were tested.  All \ensuremath{N = 5} PINGs were tested (\ensuremath{150} permutations), as well as \ensuremath{100} random \ensuremath{N = 6} PINGs.  As expected, none of the non-isomorphic problems achieved a zero energy state and thus none were classified as isomorphic.  In other words, there were no false positives. 

\section*{Discussion}
Several observations can be made from this case study.  First, the formulation of the cost function (Hamiltonian) can have a noticeable impact on quantum annealing results.  For the graph isomorphism problem, the baseline approach (embodied in Hamiltonian \ensuremath{H_1} and in Lucas \cite{LucasIsing}) blindly creates QUBO variables for every possible vertex pair, whereas the proposed Hamiltonian \ensuremath{H_2} is more parsimonious.  Variable requirements reduced from \ensuremath{N^2} to fewer than \ensuremath{N \log_2 N} (Fig.\ ~\ref{fig:figure3}) on the graph type under study, allowing larger problems to be solved (Fig.\ ~\ref{fig:figure4} and Table ~\ref{tab:tab2}).  Along with Rieffel \cite{rieffel2015case}, this is one of the first quantum annealing studies to experimentally quantify the effect of alternative Hamiltonian formulations.  One of the impacts of this observation is increased appreciation for the fact that all quantum annealing-based solvers are actually classical-quantum hybrids and that focus must be placed on effectively partitioning the processing and optimizing the classical portion.  A caveat is in order---if the classical side is made to do too much work then the quantum annealing aspect becomes trivial and of little value.  Further work is needed in identifying the specific strengths of annealing processors, and in leveraging the two sides appropriately. 

A second observation is that using majority voting error correction during post-processing can in some cases provide a benefit.  Previously, such majority voting was evaluated for a different set of problems (scheduling) and was not found to provide a significant benefit\cite{rieffel2015case}.  In our context, there were many problems for which the zero energy ground state solution was only achieved when using this post-processing; without this form of error correction (in other words, when all solutions containing a broken chain were discarded), false negatives occurred.  For instance, at \ensuremath{N = 12, 53} of \ensuremath{83} embedded problems were solved on the first job without using error correction, and an additional \ensuremath{12} problems were solved by applying error correction (Table ~\ref{tab:tab2}).  Classical error correction strategies other than majority voting should be explored and assessed in future studies, and their costs quantified.

To our knowledge, the evaluated solver is the first validated, experimental implementation of a QA-based graph isomorphism solver.  While it ultimately classified every embeddable problem correctly and demonstrated clear advantages over the baseline approach, it has serious limitations as a graph isomorphism solver.  The problem sizes are not competitive with those handled by classical solvers, which can handle \ensuremath{G\left(N, 0.5\right)} graphs with thousands of vertices \cite{mckay2014practical} and even for the hardest graph types can handle hundreds of vertices before running into difficulty \cite{McKay}.  Similarly, the scaling of the total annealing times (Fig.\ ~\ref{fig:figure5}) is not competitive with classical scaling \cite{mckay2014practical}.  Ultimately, new approaches are likely needed if quantum annealing is to contribute to graph isomorphism theory or practice.  Fortunately, the case study provides some new insight into experimental quantum annealing, and contributes methods that have relevance beyond the GI problem.  It is hoped that the experimental evaluation of alternative Hamiltonian formulations adds to the understanding of the factors affecting quantum annealing performance, and that the demonstration of majority voting raises new questions about the role of post-processing for a variety of problems.   

\section*{Methods}
{\small
Quantum annealing experiments were performed on the D-Wave Two machine housed at USC ISI and operated by the USC-Lockheed Martin Quantum Computing Center.  Experiments were conducted in October and November, 2014.  The working graph of the machine's Vesuvius-6 quantum annealing processor consisted of \ensuremath{504} qubits and \ensuremath{1427} couplers during this period.  The pattern of working qubits is shown in Figure ~\ref{fig:figure6}.  The qubit temperature was estimated to be \ensuremath{16 \pm 1} mK.  Additional processor specifications include a maximum anti-ferromagnetic mutual inductance of \ensuremath{1.33}, and \ensuremath{\frac{1}{f}} amplitude of \ensuremath{7.5 \pm 1 \frac{\mu \phi_0}{\sqrt{Hz}}}.

\begin{figure}[H]
\centering
    \includegraphics[scale=1]{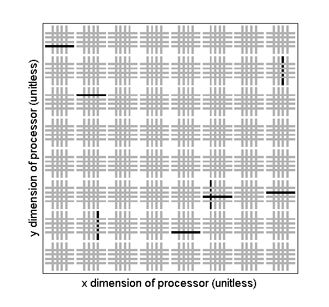}
    \caption{{\bf Physical layout of the working qubits in the USC-LM D-Wave Two Vesuvius-6 processor as of October 10, 2014.}  \ensuremath{504} working qubits (gray), \ensuremath{8} non-working (black).  All \ensuremath{1427} couplers connecting working qubits are part of the working graph.}
    \label{fig:figure6}
\end{figure}

Simple undirected \ensuremath{N}-vertex graphs were constructed according to the Erd\H{o}s-R\'{e}nyi \ensuremath{G\left(n, p\right)} model \cite{erd6s1960evolution} with \ensuremath{n = N} and with the probability \ensuremath{p} of including each edge equal to \ensuremath{0.5}.  Non-isomorphic pairs were generated by creating two graphs as above and checking for non-isomorphism using the MATLAB \texttt{graphisomorphism()} function.  Isomorphic pairs were generated by generating a single graph then applying a random permutation to arrive at the second graph.  For each pair of input graphs, an Ising model was created using equation (~\ref{eq:lucas-modified}) or (~\ref{eq:degree-based-zick}).  Programming was performed using MATLAB R2014a win64 and the D-Wave MATLAB pack 1.5.2-beta2.  The current version of the D-Wave \texttt{sapiFindEmbedding()} function cannot embed Ising models with more than one connected component (i.e. a set of variables that interact only with each other and not any of the remaining variables); therefore, models with this characteristic were not included in the input data.  When attempting to generate \ensuremath{100} input pairs for each size, such disconnected models occurred no more than \ensuremath{4} times for each size \ensuremath{N \ge 14}.  Similarly, the heuristic cannot accept models with fewer than two variables, so in the rare case of a trivial Ising problem with fewer than two variables (e.g. a non-isomorphic pair with no matching degrees), dummy variables were added to the problem.

The \ensuremath{h_i} values of the Ising problem were split evenly across each qubit in the associated chain in the embedded Ising problem.  The \ensuremath{J_{ij}} values of the Ising problem were assigned to a single coupler connecting two variable chains in the embedded problem.  The magnitudes of the embedded \ensuremath{\mathbf{h'}} and \ensuremath{\mathbf{J'}} were scaled together such that the maximum magnitude reached \ensuremath{20\%} of the full range supported by the processor; the range of the embedded \ensuremath{h_i'} values was \ensuremath{\left[-0.4, 0.4\right]} and the range of the embedded \ensuremath{J_{ij}'} values coupling different variables was \ensuremath{\left[-0.2, 0.2\right]}.  This \ensuremath{20\%} value was determined empirically to provide good performance on the median difficulty problem at the largest sizes.  Subsequently, the \ensuremath{J_{ij}'} values connecting physical qubits within a chain were set to the maximum ferromagnetic value \ensuremath{(-1)}.  A single random spin gauge transformation \cite{boixo2014evidence} was then applied to each embedded problem, with a gauge factor \ensuremath{a_i \in \left\{-1, 1\right\}} associated with each qubit and transformation \ensuremath{h_i' \rightarrow a_i h_i'; J_{ij}' \rightarrow a_i a_j J_{ij}'}.  One job was submitted to the quantum annealer per embedded problem; some Ising problems were associated with multiple embedded problems and jobs.  After each programming cycle, the processor was allowed to thermalize for \ensuremath{10} ms (the maximum supported by the machine).  The annealing time was set to the minimum value of \ensuremath{20} \ensuremath{\mu s}.  The number of annealing and readout cycles per programming cycle was \ensuremath{40000}, which allowed the total job time to be within the limits of the machine (\ensuremath{1} s).  The readout thermalisation time was set to the default value of \ensuremath{0}.  Regarding error correction through majority voting of chains of physical qubits, ties were broken by choosing the spin up state.  The probability of achieving the zero energy state on job \ensuremath{k} is denoted

\begin{align}
P_{0, k} &= \frac{\text{number of annealing cycles achieving zero energy}}{\text{number of annealing cycles}}.
\end{align}

When multiple jobs are required, we calculate the geometric mean in the style of Boixo et al. \cite{boixo2014evidence}:

\begin{align}
\bar{P_0} &= 1 - \prod^K_{k=1} \left(1 - P_{0, k}\right)^{\frac{1}{K}}.
\end{align}

The total annealing time required to reach \ensuremath{0.99} probability of success was calculated by multiplying the annealing time by the expected number of annealing cycles (repetitions \ensuremath{R}) using the formula \cite{boixo2014evidence}:

\begin{align}
R &= \ceil[\bigg]{ \frac{\ln \left(1-0.99\right)}{\ln \left(1 - \bar{P_0}\right)} }.
\end{align}
}
\pagebreak

\bibliography{graph-iso-isi}

\begin{thebibliography}{10}

\bibitem{mcgeoch2014adiabatic}
C.~C. McGeoch, {\em Adiabatic quantum computation and quantum annealing: Theory
  and practice}.
\newblock Morgan \& Claypool Publishers, 2014.

\bibitem{boixo2014evidence}
S.~Boixo, T.~F. R{\o}nnow, S.~V. Isakov, Z.~Wang, D.~Wecker, D.~A. Lidar, J.~M.
  Martinis, and M.~Troyer, ``Evidence for quantum annealing with more than one
  hundred qubits,'' {\em Nature Physics}, vol.~10, no.~3, pp.~218--224, 2014.

\bibitem{lanting2014entanglement}
T.~Lanting, A.~Przybysz, A.~Y. Smirnov, F.~Spedalieri, M.~Amin, A.~Berkley,
  R.~Harris, F.~Altomare, S.~Boixo, P.~Bunyk, {\em et~al.}, ``Entanglement in a
  quantum annealing processor,'' {\em Physical Review X}, vol.~4, no.~2,
  p.~021041, 2014.

\bibitem{Albash2015}
T.~Albash, T.~Rønnow, M.~Troyer, and D.~Lidar, ``Reexamining classical and
  quantum models for the {D-Wave One} processor,'' {\em The European Physical
  Journal Special Topics}, vol.~224, no.~1, pp.~111--129, 2015.

\bibitem{boixo2014computational}
S.~Boixo, V.~N. Smelyanskiy, A.~Shabani, S.~V. Isakov, M.~Dykman, V.~S.
  Denchev, M.~Amin, A.~Smirnov, M.~Mohseni, and H.~Neven, ``Computational role
  of collective tunneling in a quantum annealer,'' {\em arXiv preprint
  arXiv:1411.4036 [quant-ph]}, 2014.

\bibitem{mckay2014practical}
B.~D. McKay and A.~Piperno, ``Practical graph isomorphism, {II},'' {\em Journal
  of Symbolic Computation}, vol.~60, pp.~94--112, 2014.

\bibitem{babai1983canonical}
L.~Babai and E.~M. Luks, ``Canonical labeling of graphs,'' in {\em Proceedings
  of the fifteenth annual ACM symposium on Theory of computing}, pp.~171--183,
  ACM, 1983.

\bibitem{reiter2012limits}
E.~E. Reiter and C.~M. Johnson, {\em Limits of computation: an introduction to
  the undecidable and the intractable}.
\newblock CRC Press, 2012.

\bibitem{moore2008symmetric}
C.~Moore, A.~Russell, and L.~J. Schulman, ``The symmetric group defies strong
  {F}ourier sampling,'' {\em SIAM Journal on Computing}, vol.~37, no.~6,
  pp.~1842--1864, 2008.

\bibitem{hallgren2010limitations}
S.~Hallgren, C.~Moore, M.~R{\"o}tteler, A.~Russell, and P.~Sen, ``Limitations
  of quantum coset states for graph isomorphism,'' {\em Journal of the ACM
  (JACM)}, vol.~57, no.~6, p.~34, 2010.

\bibitem{kumar2009external}
Y.~Kumar and P.~Gupta, ``External memory layout vs. schematic,'' {\em ACM
  Transactions on Design Automation of Electronic Systems (TODAES)}, vol.~14,
  no.~2, p.~30, 2009.

\bibitem{king2014algorithm}
A.~D. King and C.~C. McGeoch, ``Algorithm engineering for a quantum annealing
  platform,'' {\em arXiv preprint arXiv:1410.2628 [cs.{DS}]}, 2014.

\bibitem{rieffel2015case}
E.~G. Rieffel, D.~Venturelli, B.~O'Gorman, M.~B. Do, E.~M. Prystay, and V.~N.
  Smelyanskiy, ``A case study in programming a quantum annealer for hard
  operational planning problems,'' {\em Quantum Information Processing},
  vol.~14, no.~1, pp.~1--36, 2015.

\bibitem{hen2012solving}
I.~Hen and A.~Young, ``Solving the graph-isomorphism problem with a quantum
  annealer,'' {\em Physical Review A}, vol.~86, no.~4, p.~042310, 2012.

\bibitem{vinci2014hearing}
W.~Vinci, K.~Markstr{\"o}m, S.~Boixo, A.~Roy, F.~M. Spedalieri, P.~A.
  Warburton, and S.~Severini, ``Hearing the shape of the {I}sing model with a
  programmable superconducting-flux annealer,'' {\em Scientific Reports},
  vol.~4, 2014.

\bibitem{PhysRevA.89.022342}
F.~Gaitan and L.~Clark, ``Graph isomorphism and adiabatic quantum computing,''
  {\em Phys. Rev. A}, vol.~89, p.~022342, Feb 2014.

\bibitem{LucasIsing}
A.~Lucas, ``Ising formulations of many {NP} problems,'' {\em Frontiers in
  Physics}, vol.~2, no.~5, 2014.

\bibitem{boixo2013experimental}
S.~Boixo, T.~Albash, F.~M. Spedalieri, N.~Chancellor, and D.~A. Lidar,
  ``Experimental signature of programmable quantum annealing,'' {\em Nature
  Communications}, vol.~4, 2013.

\bibitem{choi2008minor}
V.~Choi, ``Minor-embedding in adiabatic quantum computation: {I}. {T}he
  parameter setting problem,'' {\em Quantum Information Processing}, vol.~7,
  no.~5, pp.~193--209, 2008.

\bibitem{erd6s1960evolution}
P.~Erd\H{o}s and A.~R{\'e}nyi, ``On the evolution of random graphs,'' {\em
  Publ. Math. Inst. Hungar. Acad. Sci}, vol.~5, pp.~17--61, 1960.

\bibitem{cai2014practical}
J.~Cai, W.~G. Macready, and A.~Roy, ``A practical heuristic for finding graph
  minors,'' {\em arXiv preprint arXiv:1406.2741 [quant-ph]}, 2014.

\bibitem{McKay}
B.~D. McKay, ``Graph isomorphism,'' in {\em Handbook of Graph Theory}
  (P.~Zhang, ed.), Chapman and Hall/CRC, 2013.

\end{thebibliography}
\bibliographystyle{ieeetr}

\section*{Acknowledgements}
We would like to thank Itay Hen for helpful discussions and suggestions, and Federico Spedalieri for feedback on an early version of the manuscript.  O.S. would like to thank Professor Samuel J. Lomonaco, Jr. for insight into the graph isomorphism problem.  This material is based in part upon work supported by the Defense Advanced Research Projects Agency (DARPA) under Contract No. HR001-11-C-0041.  The views expressed are those of the authors and do not reflect the official policy or position of the Department of Defense or the U.S. Government.  Distribution Statement “A” (Approved for Public Release, Distribution Unlimited).  

\section*{Author Contributions}
K.Z. and O.S. conceived the research; O.S. performed mathematical modeling and developed an algorithm for generating an Ising model; K.Z. designed and implemented the solvers, conducted experiments, and wrote the manuscript text; M.F. and K.Z. supervised the research; M.F. reviewed the manuscript and contributed revisions.

\end{document}